\documentclass[11pt, twocolumn,trackchanges]{aastex631}
\usepackage{xcolor}
\usepackage{CJK}
\usepackage{bm}
\usepackage{soul}
\definecolor{victoria}{rgb}{0.76, 0.36, 0.42}

\usepackage{float}
\usepackage{amsmath}
\usepackage{pgfplots}
\usepackage{units}
\usepackage{multirow}
\pgfplotsset{compat=1.15}
\newcommand\bluesout{\bgroup\markoverwith{\textcolor{blue}{\rule[0.5ex]{2pt}{0.4pt}}}\ULon}

\date{\today}

\begin{document}
\begin{CJK*}{UTF8}{gbsn}

\title{Low-frequency Intermittency and Structures in the Solar Wind at 1 au}

\author[0009-0008-8723-610X]{Jiaming Wang (王嘉明)}
\affiliation{Department of Physics and Astronomy, University of Delaware}

\author[0000-0003-4168-590X]{Francesco Pecora}
\affiliation{Department of Physics and Astronomy, University of Delaware}

\author[0000-0003-2965-7906]{Yan Yang (杨艳)}
\affiliation{Department of Physics and Astronomy, University of Delaware}

\author[0000-0001-7224-6024]{William H. Matthaeus}
\affiliation{Department of Physics and Astronomy, University of Delaware}

\begin{abstract}

Intermittency in the solar wind is commonly studied within the inertial and dissipative ranges, where scale-dependent magnetic field distributions become increasingly non-Gaussian toward smaller scales until dissipation becomes important. Conversely, whether Gaussianity is recovered at large scales remains unclear. To address this, we systematically examine magnetic field increment kurtosis over scales from 1 minute to 1 year using more than two decades of in situ observations from NASA's Wind and ACE spacecraft. We find that although kurtosis tends toward Gaussian value of 3 near the correlation scales, it generally remains elevated (super-Gaussian) at larger scales. Kurtosis also varies substantially over time, with pronounced super-Gaussian intervals during high solar activity phases and sporadic sub-Gaussian intervals primarily in the radial component. These results provide evidence for large-scale intermittency in the solar wind, potentially arising from mixing of different solar wind streams and nonstationary driving of solar sources.
\end{abstract}

\section{Introduction}

Intermittency is a well-known property of turbulence in which events with extreme values occur with enhanced likelihood relative to standard reference random processes~\citep{Novikov71, Frisch95}. 
In classical turbulence theory, intermittency is interpreted as localized bursts of energy dissipation~\citep{Obukhov62, Kolmogorov62} and commonly manifests as elevated probability of large fluctuations in the primitive variables, such as component increments of the velocity or, in plasmas, of the magnetic field. The corresponding probability distributions (PDFs) therefore develop enhanced ``tails'' and sharp central peaks~\citep{matthaeus2015intermittency}. Such features of the PDF necessarily influence moments of all orders, leading to formal theories such as the multifractal scaling of high-order moments~\citep{Frisch95}. A compact and readily computable quantity to detect intermittency in a scalar field is the normalized fourth-order moment, or kurtosis $K$. In particular, a Gaussian PDF has $K=3$, whereas $K>3$ is commonly associated with intermittency. 

The solar wind exhibits turbulent intermittency in the magnetohydrodynamic (MHD) inertial range~\citep[see, e.g.,][]{Horbury97, SorrisoValvo99, Wan09, Wan12}, in analogy to the hydrodynamic (HD) intermittency~\citep{Anselmet84, She94}. As in HD, much attention is devoted to understanding intermittency variation across the inertial range, where kurtosis generally increases toward smaller dissipative scales~\citep{Sreenivasan97}. This trend also reconciles with the emergence of intense coherent structures occupying localized regions, also referred to as intermittent structures, with current sheets being one example~\citep{Matthaeus15}. The behavior at large scales, however, is less well established. One might expect, in a large-scale homogeneous medium, that intermittency would diminish and fluctuations become progressively more Gaussian at scales comparable to or greater than the correlation scales~\citep{Frisch91, Kailasnath92}. Do large-scale solar wind signals ever reach Gaussianity? Early solar wind observations suggest an approach to Gaussianity ~\citep{SorrisoValvo99} but not a strict convergence and therefore 
motivate a systematic investigation over a broader range of scales and times.

Several processes may prevent long-wavelength, low-frequency solar wind signals from approaching Gaussian states devoid of intermittency: large coherent structures and transient events (e.g., coronal mass ejections and heliospheric current sheet crossings) produce intense fluctuations; long-memory dynamics related to the $1/f$ range may also contribute to large-scale non-Gaussianity~\citep{Bruno13}. Such low-frequency intermittency is anticipated in the classical paper by \citet{Obukhov62} in the context of high-altitude atmospheric winds, where slow variations in large-scale environments modulate the local energy dissipation rate, leading to non-Gaussian statistics through mixing of different regimes. 
In the solar wind, analogous long-memory variations may reflect coronal organization and indirect statistical connections to the solar dynamo~\citep{Wang24_1overf}. 

Here, we use the kurtosis of magnetic field increment distributions to quantify intermittency and deviations from Gaussian statistics in the 1 au solar wind. Taking advantage of multi-decadal in situ observations, we examine scale-dependent kurtosis from daily to monthly scales and their temporal variability. The paper is organized as follows: Section~\ref{sec:data} describes the datasets and kurtosis estimations, Section~\ref{sec:result} presents results on the temporal and scale-dependent variations of kurtosis, and Section~\ref{sec:discussion} summarizes and discusses
the results.

\section{Data and analysis procedure}
\label{sec:data}

We use magnetic field measurements from NASA's Wind and Advanced Composition Explorer (ACE) spacecraft located at the L1 Lagrange point. The Wind observations span 2003-01-01 to 2025-12-31, and the ACE observations span 1998-02-05 to 2024-07-10. Magnetic field vectors are collected from the Magnetic Field Investigation instrument onboard Wind~\citep[MFI;][]{Lepping95} and the Magnetometer onboard ACE~\citep[MAG;][]{Smith98}. Wind/MFI vectors at 1-minute cadence in the radial-tangential-normal (RTN) coordinates are directly accessible from a web portal \citet{Koval23}.
Similarly, ACE/MAG vectors at 1-second resolution in RTN coordinates are accessed from \citet{Smith22_l2}, and
then re-sampled at 1-minute cadence.

\begin{figure*}
\centering
    \includegraphics[angle=0,width=\columnwidth]{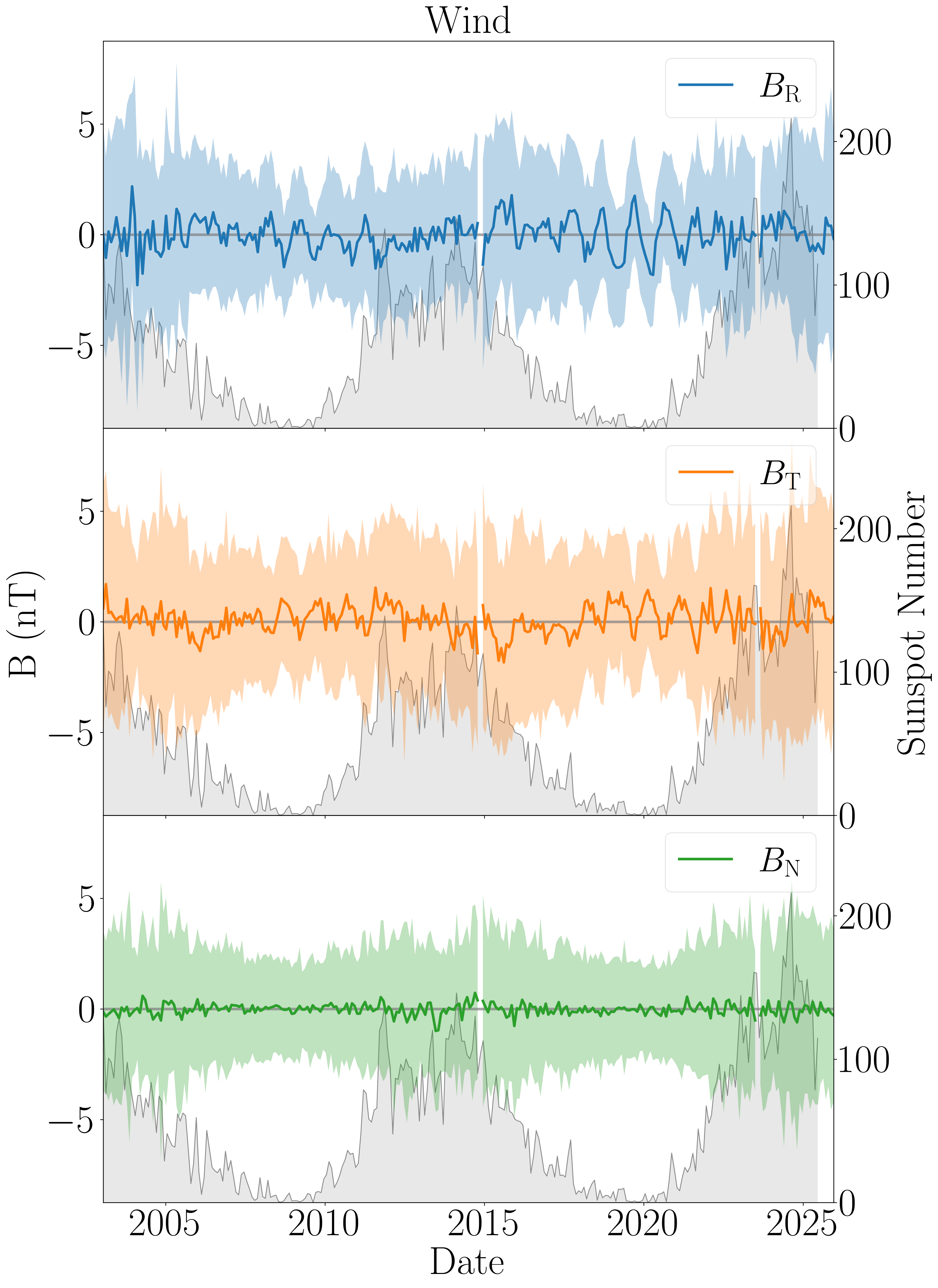}
    \includegraphics[angle=0,width=\columnwidth]{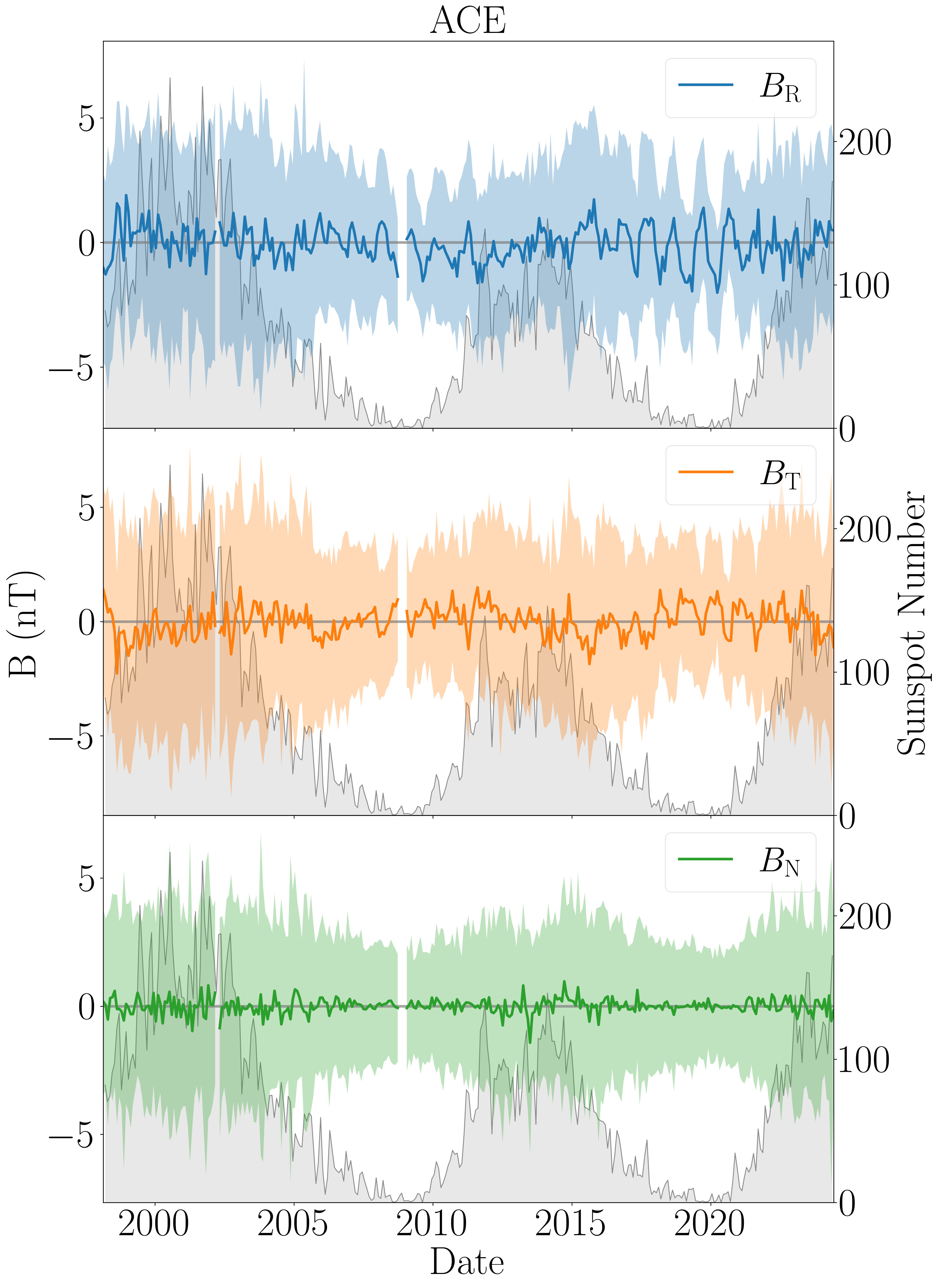}
    \caption{Magnetic field component time series as observed by Wind (left) and ACE (right). Solid lines and shaded regions represent 30-day averages $\pm$ one standard deviation. Gray background shows sunspot number over time.}
\label{fig:Bts}
\end{figure*}

Both datasets are treated as continuous time series containing occasional gaps, arising primarily from temporary instrument outages (Fig.~\ref{fig:Bts}). A quality-control procedure is applied throughout each dataset to remove anomalous measurements -- individual data points falling beyond 3 standard deviations of the local 3-hour running mean are identified as outliers and filled with NaN values (left as a gap; not interpolated). We show 30-day averages of magnetic field components and their standard deviations in Fig.~\ref{fig:Bts}.

For each magnetic field component $B_i$, $i = \mathrm{R}, \mathrm{T}, \mathrm{N}$, we define the increment at lag $\tau$ measured at time $t$ as $\delta B_i(t, \tau) = B_i(t+\tau) - B_i(t)$. The scale-dependent increment kurtosis $K_i = \langle \delta B_i^4\rangle/ \langle \delta B_i^2\rangle^2$ is estimated using two methods: The direct ensemble estimate is
\begin{equation}
    K_i^{\mathrm{ens}}(\tau) = N \frac{\sum_N \delta B_i^4(\tau)}{ \left[ \sum_N \delta B_i^2(\tau) \right]^2},
\label{eq:kens}
\end{equation}
where $N$ is the number of valid increments in time. We also compute the kurtosis from an estimated increment probability distribution $P(\delta B_i(\tau))$:
\begin{equation}
    K_i^{\mathrm{PDF}} (\tau) = \frac{\int_{-\infty}^\infty \delta B_i(\tau)^4 P(\delta B_i(\tau)) d \delta B_i(\tau)}{\left[ \int_{-\infty}^\infty \delta B_i(\tau)^2 P(\delta B_i(\tau)) d \delta B_i(\tau) \right]^2},
\label{eq:kpdf}
\end{equation}
where very large $\delta B_i$ values are rare enough to assume convergence of the integral. The distribution $P(\delta B_i(\tau))$ is constructed using equal-count binning, with approximately 1000 counts per bin and bin centers defined as the mean value within each bin. This procedure slightly underestimates the kurtosis due to reduced weighting of extreme tail events, as compared to the ensemble estimate~\citep{Breech03}.

We adopt $K_i^{\mathrm{ens}}$ as the primary measured quantity and use $K_i^{\mathrm{PDF}}$ as a consistency check. The two quantities are not expected to agree exactly, but they should exhibit similar scale- and time-dependent variations.

\section{Results}
\label{sec:result}

\begin{figure*}
\centering
    \includegraphics[angle=0,width=\columnwidth]{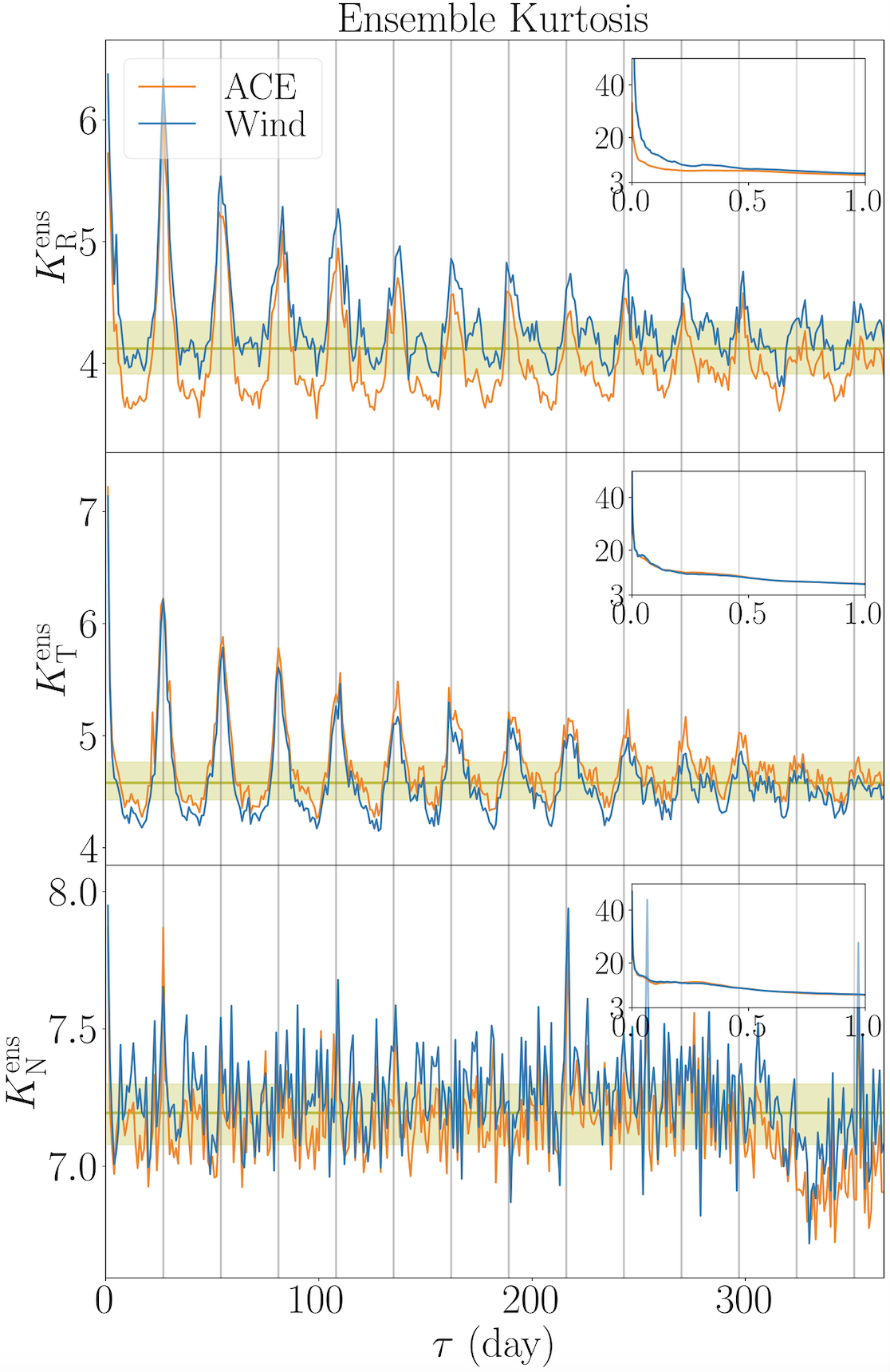}
    \includegraphics[angle=0,width=\columnwidth]{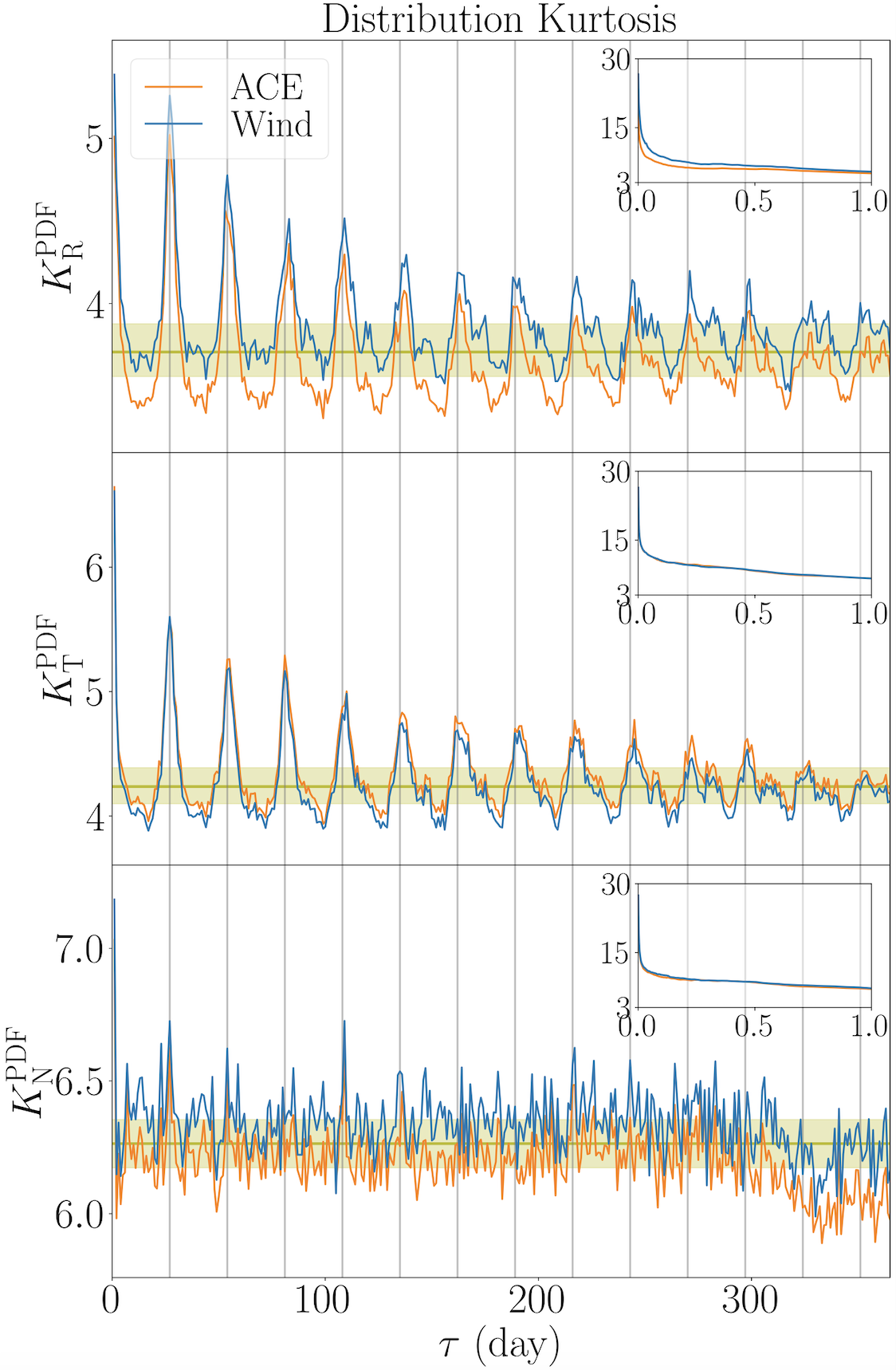}
    \caption{Scale-dependent ensemble (left) and distribution (right) kurtosis of magnetic field increments from Wind (blue) and ACE (orange). Main panels show scales from 1 day to 1 year; insets show sub-day scales. Horizontal green lines represent the 50th percentile of kurtosis. Boundaries of shaded regions represent the 25th and 75th percentiles.}
\label{fig:sdk}
\end{figure*}

We show in Fig.~\ref{fig:sdk} the scale-dependent kurtosis, $K_i^\mathrm{ens}(\tau)$ and $K_i^\mathrm{PDF}(\tau)$ as defined in Eqs.~\ref{eq:kens} and~\ref{eq:kpdf}, computed using WIND and ACE observations over their overlapping period (2003-01-01 to 2023-12-31) that spans almost one complete magnetic solar cycle. The kurtosis estimates from both methods show good agreement between Wind and ACE, albeit with slightly higher Wind kurtosis in the $\mathrm{R}$ and $\mathrm{N}$ directions and higher ACE kurtosis in the $\mathrm{T}$ direction. The close correspondence between Wind and ACE in the scale-dependent variations provides confidence in fourth-order moment analysis using in situ observations. For succinctness, we show only Wind data analyses for the rest of this paper.

Perhaps the most surprising result in Fig.~\ref{fig:sdk} is the persistently super-Gaussian kurtosis\footnote{While super- or sub-Gaussian have rigorous mathematical definitions, here we use these terms descriptively to refer to distributions with $K>3$ or $K<3$.} ($K>3$) across scales from 1 minute to 1 year. The insets cover scales from 1 minute to 1 day, encompassing the nominal turbulence correlation scales of around one to several hours, defined here as the $1/e$ decay scale of the magnetic field correlation trace~\citep[see][]{Ruiz14, Isaacs15, Wang26_superposition}. The main panels extend to much larger scales to probe monthly to yearly variability. Kurtosis values in all three components decrease with increasing scales within the inertial range and the larger energy-containing range but never reach 3.

We find a 27-day periodicity possibly associated with the Carrington rotation in $K_\mathrm{R}$ and $K_\mathrm{T}$ but not in $K_\mathrm{N}$, closely mirroring the periodic behavior in the second-order scale-dependent correlation functions reported in \citet{Wang26_correlation}. Because the second-order structure function $S_2(\tau)$ is related to the autocorrelation\footnote{The autocorrelation of magnetic field component can be computed as $R(\tau) = \langle B_i(t) B_i(t+\tau)\rangle - \langle B_i(t)\rangle \langle B_i(t+\tau)\rangle$; see \citet{Blackman58, Matthaeus82-convergence}.} $R(\tau)$ through $S_2(\tau) = 2[R(0) - R(\tau)]$, enhanced correlation at the 27-day scale 
can lead to enhanced kurtosis $K(\tau) = S_4(\tau) / S_2^2(\tau)$, assuming weak variation in $S_4(\tau)$. To examine this possibility, we apply a 1-day high-pass filter to 21 years of Wind data (as used in Fig.~\ref{fig:sdk}) to suppress daily-scale second-order structure and repeat the scale-dependent kurtosis analysis. The results are shown in Fig.~\ref{fig:sdk_highfilter} and the details of the filtering procedure are documented in Appendix~\ref{sec:app1}.

\begin{figure}
\centering
    \includegraphics[angle=0,width=\columnwidth]{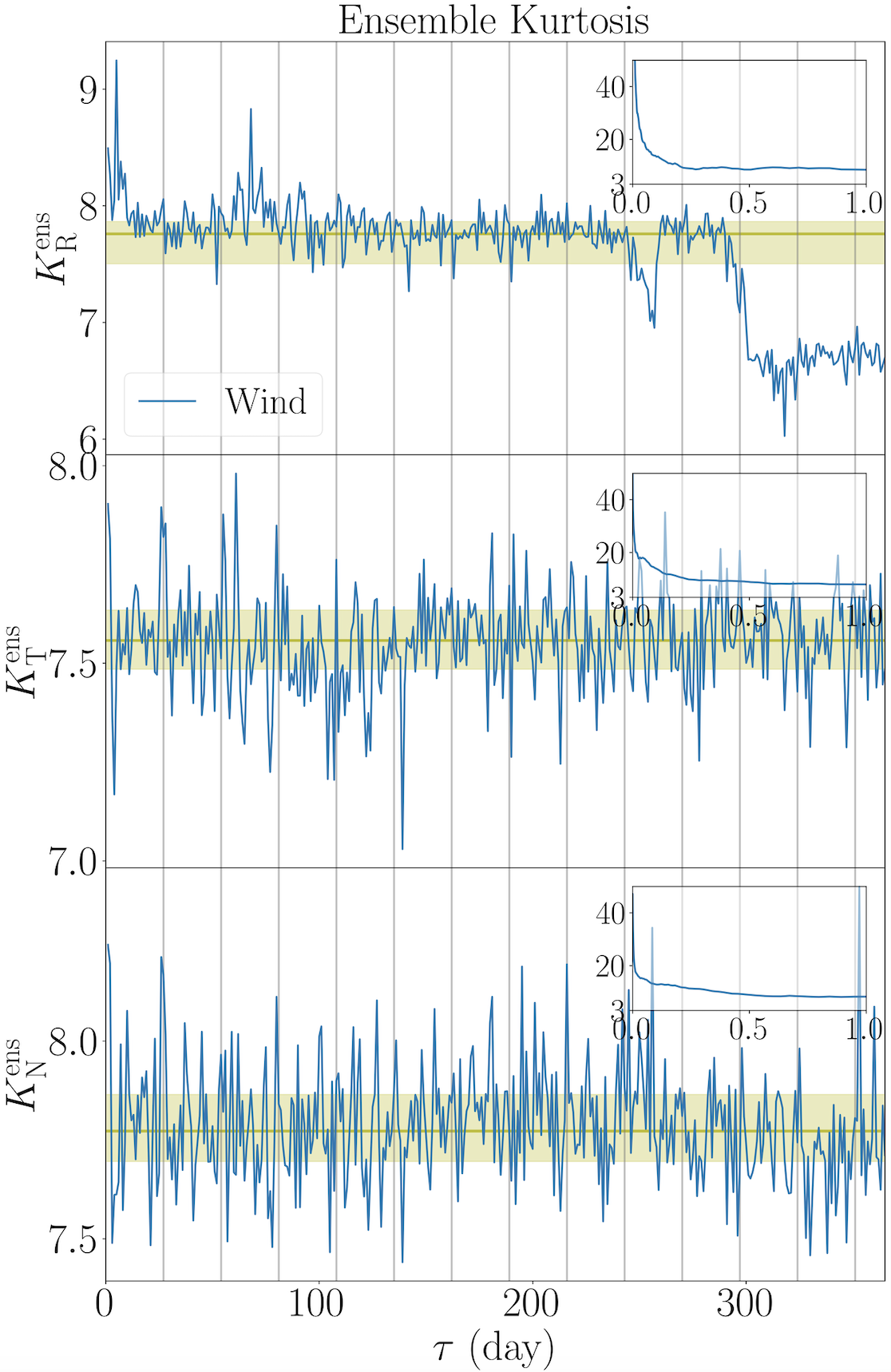}
    \caption{Scale-dependent ensemble kurtosis $K^\mathrm{ens}$ of magnetic field increments from Wind data after 1-day high-pass filtering. Main panels show scales from 1 day to 1 year; insets show sub-day scales. Horizontal green lines represent the 50th percentile of kurtosis. Boundaries of shaded regions represent the 25th and 75th percentiles.}
\label{fig:sdk_highfilter}
\end{figure}

After high-pass filtering, kurtosis remains substantially greater than 3 across scales from 1 day to 1 year. However, the 27-day periodicity disappears, indicating that it is primarily associated with second-order statistics (e.g. variations in the fluctuation energy) rather than the raw fourth-order statistics. The overall trend in the $\mathrm{N}$ component $K^\mathrm{ens}_\mathrm{N}$ remains largely unchanged before and after high-pass filtering, whereas $K^\mathrm{ens}_\mathrm{R}$ and $K^\mathrm{ens}_\mathrm{T}$ become similar to $K^\mathrm{ens}_\mathrm{N}$ after filtering. The kurtosis at scales below the filtering window (1 day) is shown in the insets, and is almost identical to that obtained from the unfiltered data (in the insets of Fig.~\ref{fig:sdk}), as expected.

The observed high kurtosis may arise from rare, bursty events, such as interplanetary coronal mass ejections and heliospheric current sheet crossings, or from slow temporal modulation of the fluctuation energy. The latter can generate non-Gaussian statistics through variance- or mean-variance-mixture models, such as the Castaing~\citep{Castaing90} and Normal Inverse Gaussian~\citep{BarndorffNielsen97, BarndorffNielsen04} models. 
To reduce contributions from annual and long-range variance mixing, in the following analysis, we compute kurtosis within 6-month windows, which provides sufficient sampling for estimating kurtosis up to 2-month scales.

\begin{figure*}
\centering
    \includegraphics[angle=0,width=\columnwidth]{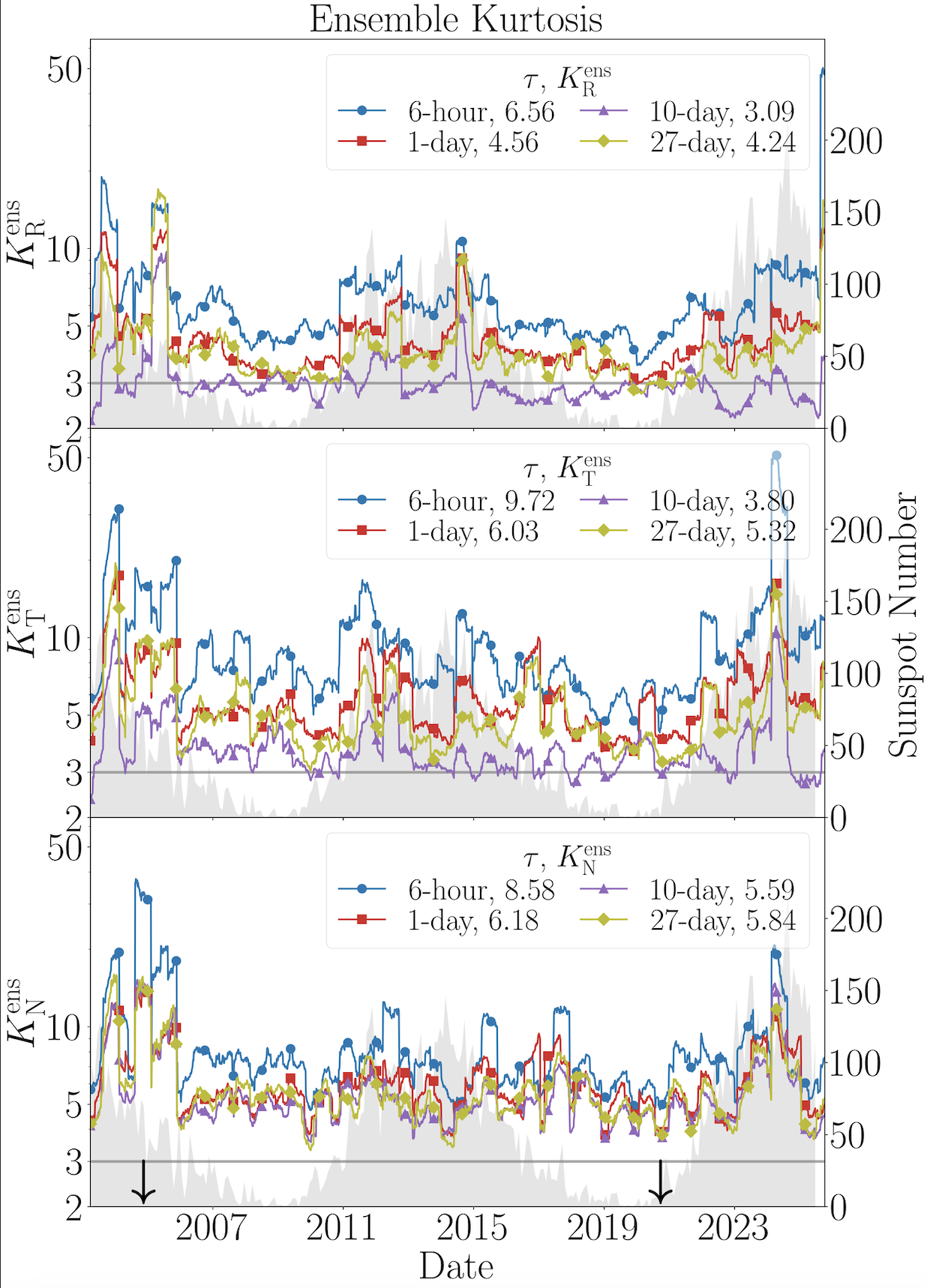}
    \includegraphics[angle=0,width=\columnwidth]{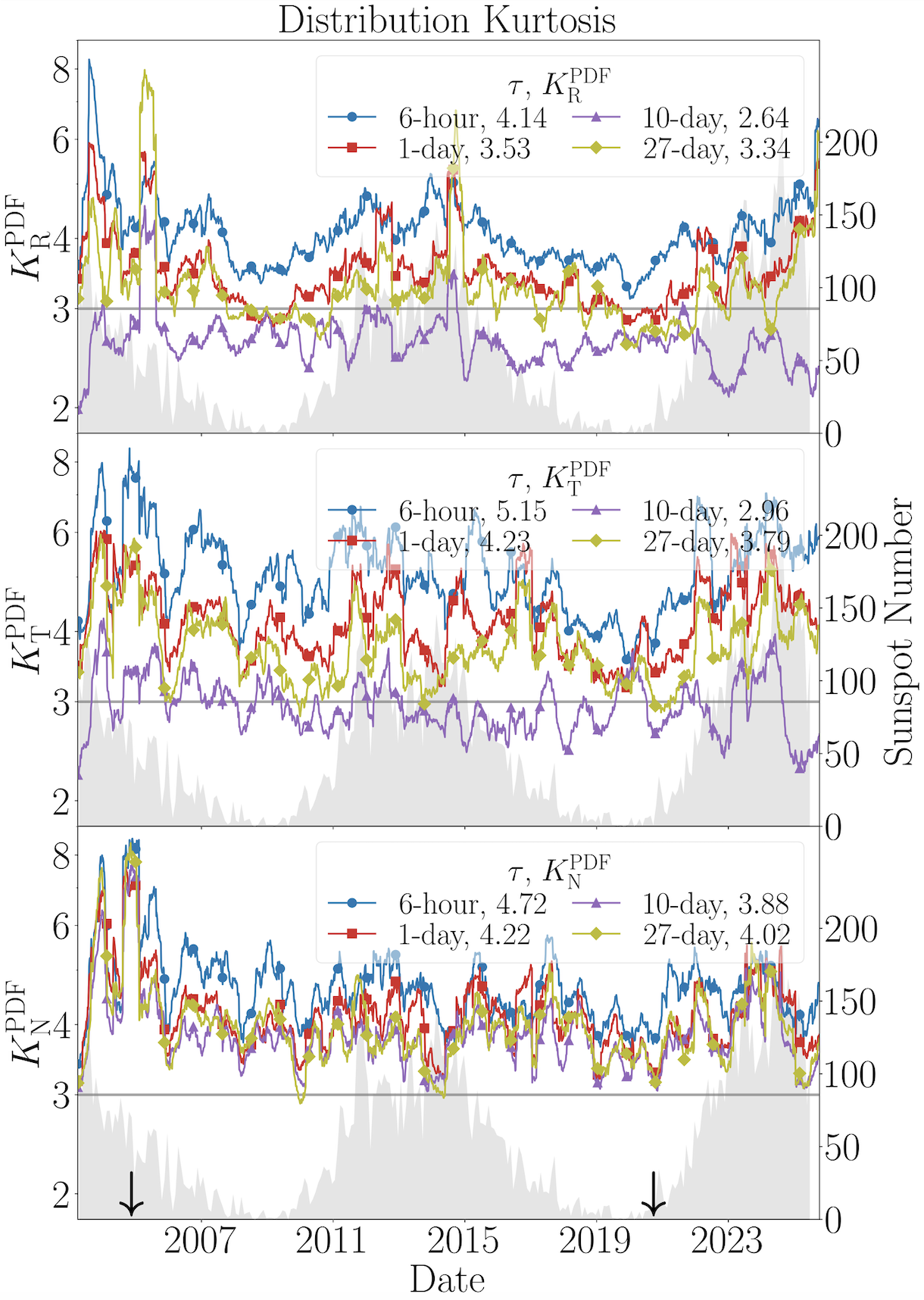}
    \caption{Time series of ensemble (left) and distribution (right) kurtosis at 6-hour (blue; circle), 1-day (red; square), 10-day (purple; triangle), and 27-day (yellow; diamond) scales from Wind. Kurtosis at each scale averaged over 23 years is listed in the legend. Arrows mark dates of high and low kurtosis with distributions shown in Fig.~\ref{fig:dist}. Gray background shows sunspot number over time.}
\label{fig:kts}
\end{figure*}

Fig.~\ref{fig:kts} shows the temporal evolution of increment kurtosis at 6-hour, 1-day, 10-day, and 27-day scales. Each kurtosis estimate uses a 6-month data window centered on time points spaced 4 days apart. On average, kurtosis is the highest at the 6-hour scale bordering the inertial range, whereas at the 10-day scale, $K^\mathrm{ens}_\mathrm{R}$ and $K^\mathrm{ens}_\mathrm{T}$ occasionally become sub-Gaussian. The $\mathrm{N}$ component noticeably maintains super-Gaussian kurtosis across the entire analyzed duration.

\begin{figure*}
\centering
    \includegraphics[angle=0,width=0.8\columnwidth]{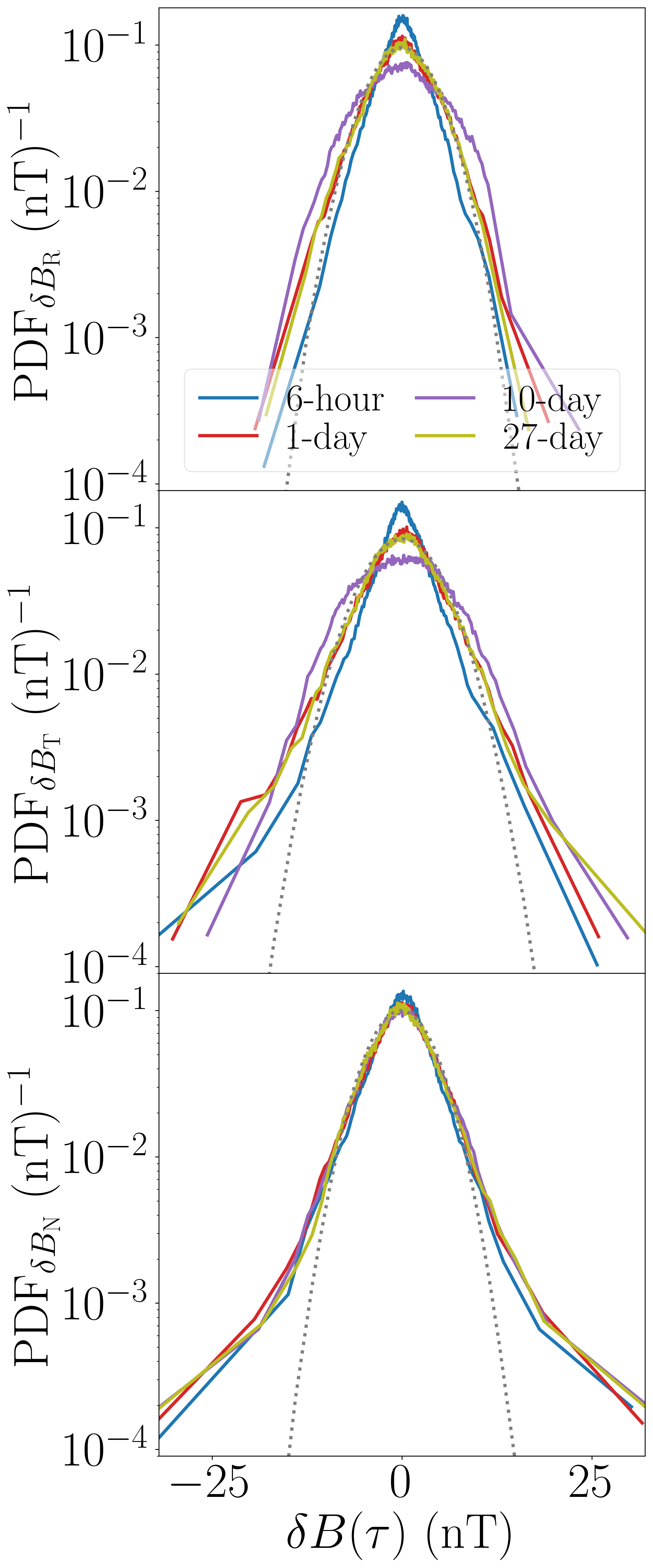}
    \includegraphics[angle=0,width=0.8\columnwidth]{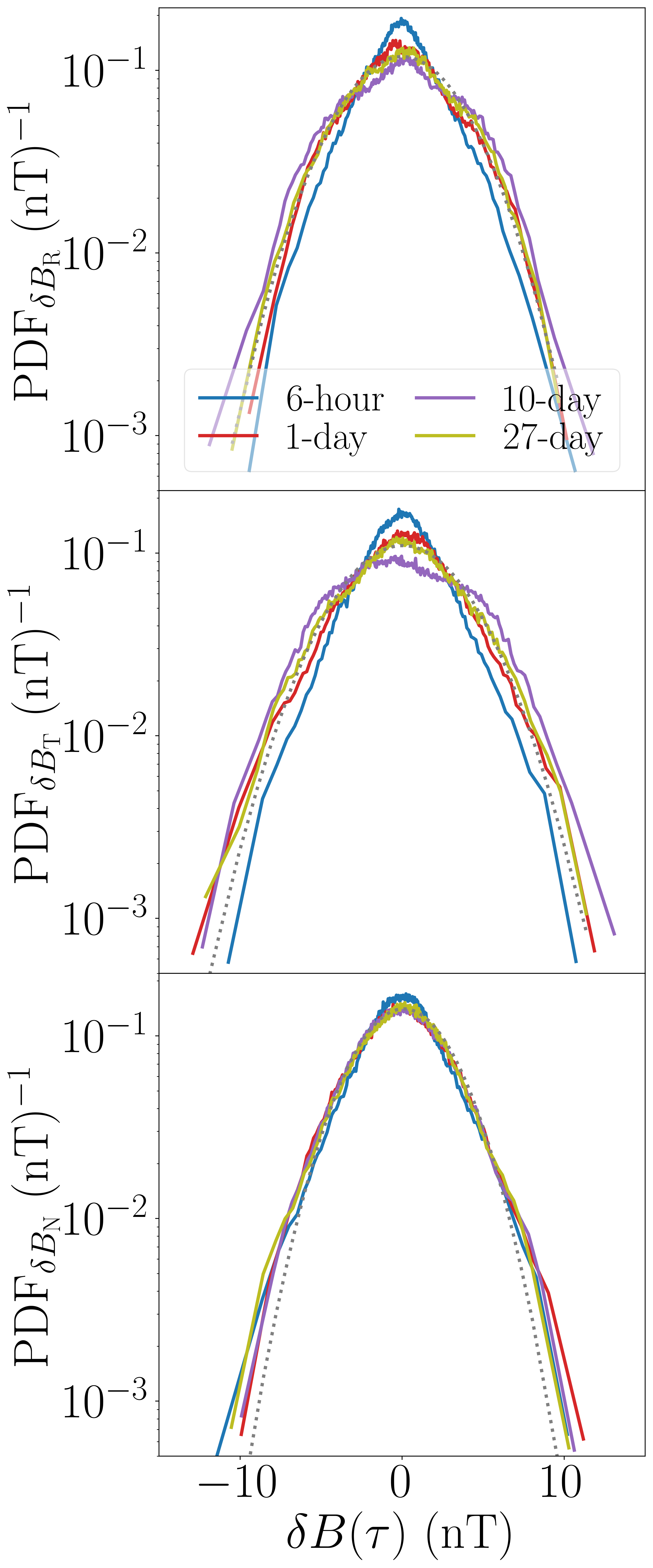}
    \caption{Distributions of magnetic field increments at 6-hour (blue), 1-day (red), 10-day (purple), and 27-day (yellow) scales from 6-month Wind data centered on date 2004-11-09 (left) and 2020-09-14 (right). Gray dotted curves show reference Gaussian distributions.}
\label{fig:dist}
\end{figure*}

Kurtosis varies substantially over time, suggesting that at any fixed scale, no single distribution can adequately characterize solar wind magnetic field increments, as attempted in numerous previous studies~\citep[e.g.,][]{SorrisoValvo99, Padhye01, SorrisoValvo15}. In particular, elevated kurtosis occurs more frequently during solar maximum phases. In Fig.~\ref{fig:dist}, we compare the shapes of increment distributions on dates 2004-11-09 and 2020-09-14, with relatively high and low kurtosis, respectively. The distributions are computed with 6-month data samples centered on these dates. As anticipated, the distributions deviate from the reference Gaussians and have sharp central peaks and heavy tails on 2004-11-09, and are close to Gaussian on 2020-09-14. We will present modeling of the distribution shapes in a subsequent paper.

\begin{figure}
\centering
    \includegraphics[angle=0,width=\columnwidth]{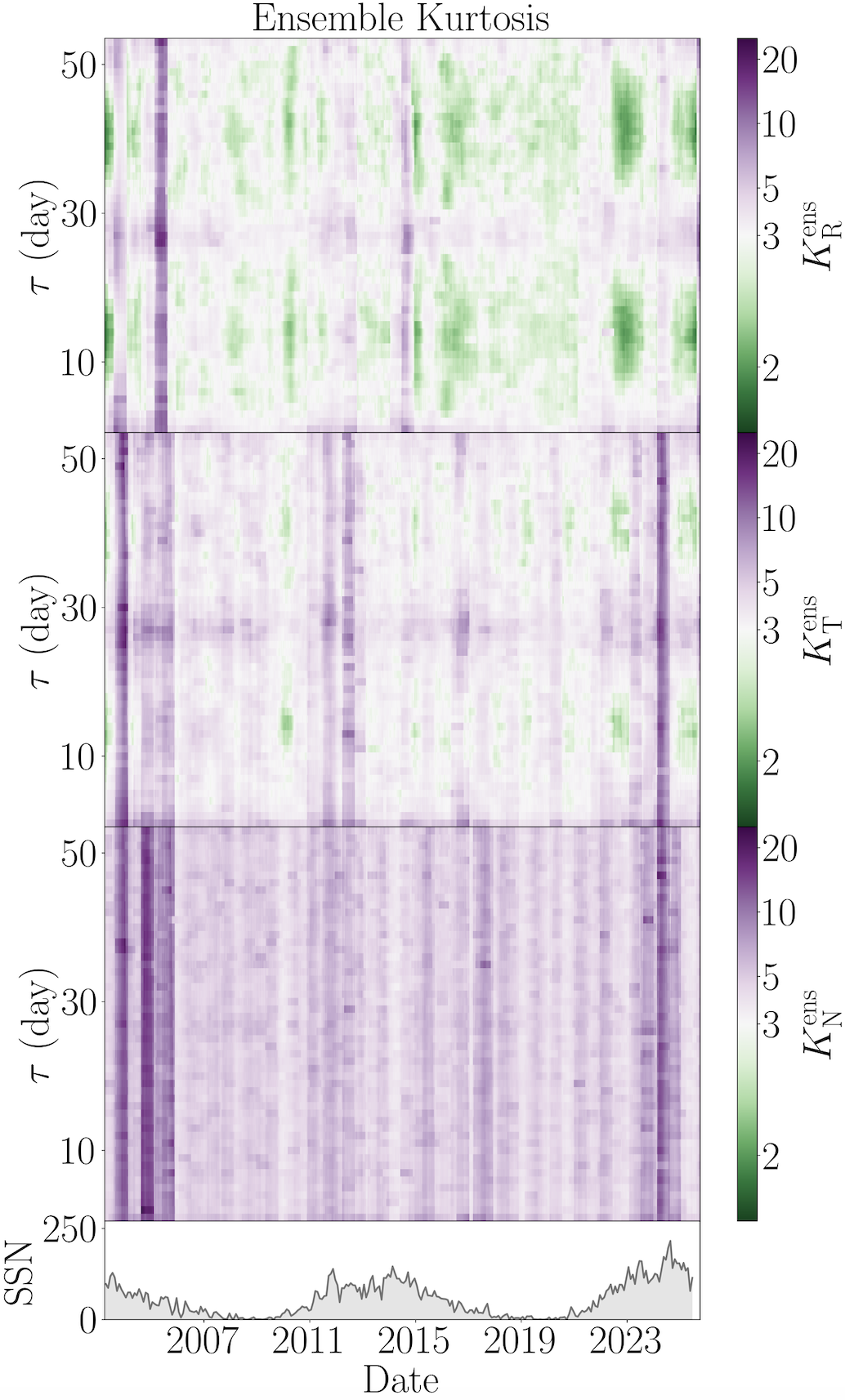}
    \caption{Colormap of scale-dependent ensemble kurtosis of magnetic field increments over time. Green represents $K < 3$ values with linear-scale intensity; purple represents $K > 3$ values with log-scale intensity. Bottom panel shows sunspot number over time.}
\label{fig:kcm}
\end{figure}

For completeness, we show the temporal evolution of $K^\mathrm{ens}$ across scales from 1 to 54 days in Fig.~\ref{fig:kcm}. The colormaps of $K^\mathrm{ens}$ reveal extended super-Gaussian periods (in purple with log-scale intensity) interspersed with localized sub-Gaussian regions (in green with linear-scale intensity), particularly in the $\mathrm{R}$ component. Super-Gaussian $K_\mathrm{T}$ and $K_\mathrm{N}$ are most evident during high solar activity phases, although $K_\mathrm{R}$ shows clear sub-Gaussian patches during the last solar maximum. 

\begin{figure}
\centering
\includegraphics[angle=0,width=\columnwidth]{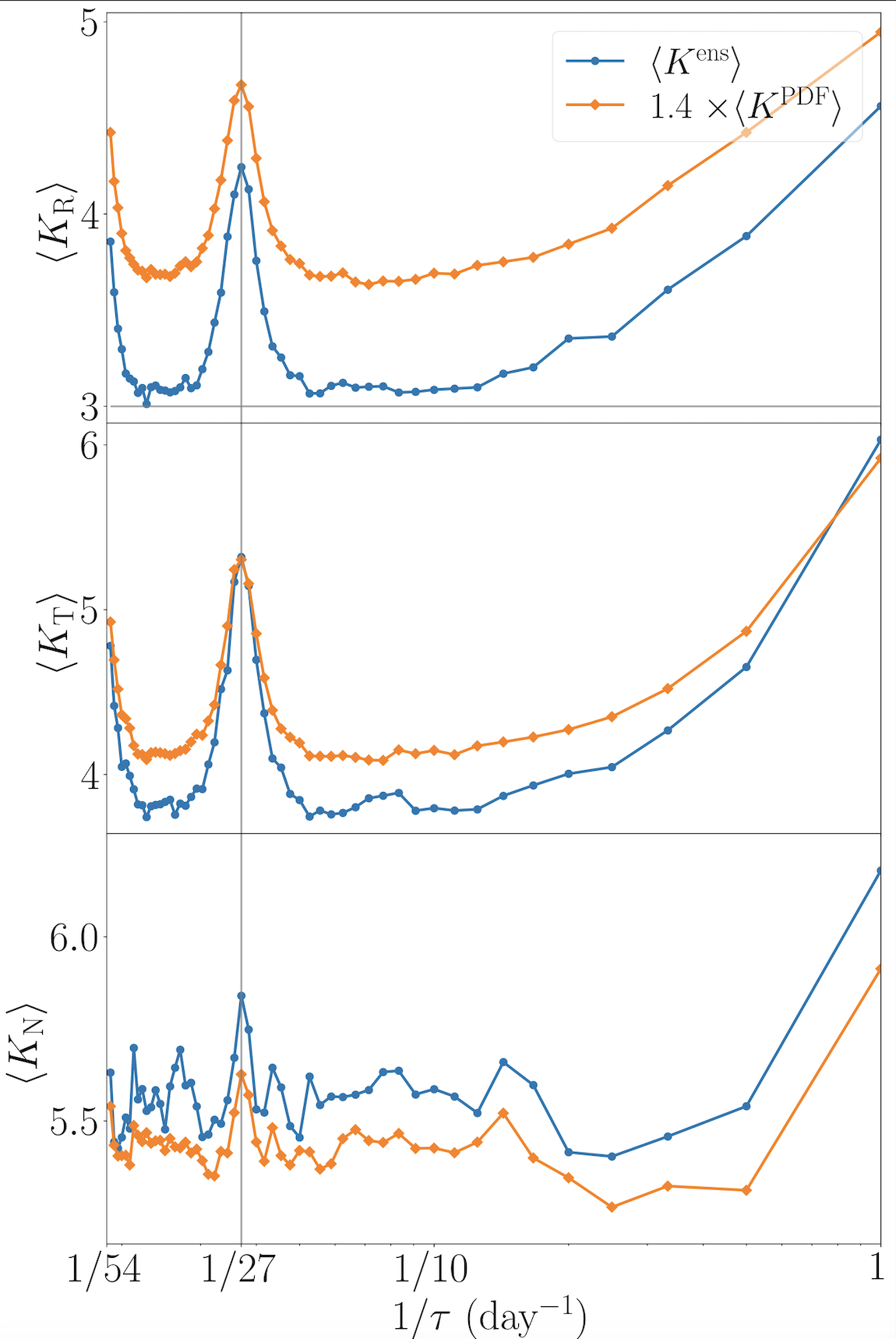}
    \caption{Scale-dependent ensemble (blue; circle) and distribution (orange; diamond) kurtosis of magnetic field increments computed with rolling 6-month intervals of Wind data, then averaged over time.}
\label{fig:avgk}
\end{figure}

Finally, Fig.~\ref{fig:avgk} shows the increment kurtosis at each lag averaged over time (i.e., a horizontal average in Fig.~\ref{fig:kcm}), here denoted by $\langle K \rangle$. The ensemble and distribution kurtosis estimates display roughly similar scale dependence. We emphasize that Fig.~\ref{fig:avgk} differs from Fig.~\ref{fig:sdk} as normalized kurtosis is not additive. Nevertheless, both reveal pronounced super-Gaussianity at 27- and 54-day scales in the $\mathrm{R}$ and $\mathrm{T}$ components. Remarkably, $\langle K_\mathrm{N} \rangle$ also has a weak local maximum at 27 days. Beyond around 10 days, excluding periodic enhancements, $\langle K_\mathrm{R} \rangle$ tends to average toward Gaussian values over 23 years, and $\langle K_\mathrm{T} \rangle$ exhibits a similar trend toward a baseline of around 3.7, while $\langle K_\mathrm{N}\rangle$ fluctuates around 5.6.

\section{Discussion}
\label{sec:discussion}

While the conventional 
picture that increment kurtosis increases toward shorter lags or higher frequencies~\citep{Frisch95} is, by and large, true, the converse -- that kurtosis at very large lags should return to the non-intermittent Gaussian value of 3 -- does not hold in the solar wind. We find that there is a tendency for kurtosis to decrease toward near-Gaussian values as the lag approaches the correlation scales, but it does not consistently reach 3. At daily to monthly lags, pronounced super-Gaussianity emerges at integer multiples of 27 days in the $\mathrm{R}$ and $\mathrm{T}$ components, whereas the $\mathrm{N}$ component remains persistently super-Gaussian without periodic structure. Kurtosis also varies considerably over time, with extended super-Gaussian intervals particularly during high solar activity phases, interspersed with sporadic sub-Gaussian patches especially in the $\mathrm{R}$ component.

A fundamental distinction between inertial range and large-scale intermittency is that the former is commonly attributed to multiplicative turbulence cascade with nonuniform local energy transfer~\citep{Kolmogorov62, Castaing90, Benzi91, Kailasnath92, She94}, whereas the latter is inherited from nonstationary driving independent of scale-local cascade. Obukhov's description of atmospheric winds~\citep{Obukhov62} connects the two narratives: variability in the sources or drivers is reflected in both large-scale non-Gaussianity and, through statistical mixing, intermittent statistics in the downstream cascade. 
In the solar wind, analogous variability could be associated with evolving coronal structures, recurrent wind streams, or magnetic activities in the chromospheric or photospheric source regions. A connection to the internal dynamics of the Sun, and therefore the dynamo, appears to be plausible, though underlying pathways remain to be investigated. Future study along these lines may provide valuable links between internal solar dynamics, predictability, and space weather. 

Similar ``solar-origin'' statements also apply to the observed heliospheric low-frequency $1/f$ spectrum~\citep{Wang24_1overf, Wang26_correlation}, implying a possible interesting relationship between the second- and fourth-order statistics of the interplanetary magnetic field. The enhanced kurtosis at 27- and 54-day lags in the $\mathrm{R}$ and $\mathrm{T}$ components is likely linked to solar-rotation-scale structures appearing in the second- rather than the raw fourth-order statistics, as indicated by the high-pass filtering test. But the persistence of elevated kurtosis across daily to monthly scales after filtering points to additional large-scale non-Gaussian sources. In addition, the tendency for kurtosis to increase after high-pass filtering is qualitatively reminiscent of the classical interpretation of HD intermittency in the inertial range~\citep{Frisch95}. Detailed verifications in MHD simulations and solar wind measurements are conducted in \citet{Wan09, Wan10, Wan12_intermittency}. An open question is whether solar wind large-scale non-Gaussianity shares statistical characteristics with inertial range turbulence.

Finally, the persistence of large-scale super-Gaussianity after temporal averaging disfavors isolated extreme events as the sole explanation for high kurtosis amplitude. Distinguishing structured modulation from rare-event contributions will require conditioned analyses, which we plan to pursue in a follow-up study.

\section*{Acknowledgements}

This research is partially supported by
the U.S. National Science Foundation, award PHY-2108834, through the NSF/DOE Partnership in Basic Plasma Science and Engineering, by the NASA IMAP project at UD under subcontract SUB0000317 from Princeton University, by the NASA/SWRI PUNCH subcontract N99054DS at the University of Delaware, and by the NASA HSR grant 80NSSC25K7757. This work is partially supported by NSF SHINE 2501387 at the University of Delaware.

\clearpage
\appendix
\section{1-day high-pass filter procedure}
\label{sec:app1}

Here we describe the 1-day high-pass filtering procedure. The high-pass-filtered signal is obtained by subtracting a low-pass-filtered version from the original signal. We first describe the mathematical formalism in the continuous, unbounded domain then proceed to the finite, discrete implementation.

For a continuous signal $x(t)$, the goal is to apply a boxcar band-limit function in the frequency space so that no energy exists above the Nyquist frequency $\omega_\mathrm{N} = \pi/M$, where $M$ is half of the filtering timescale. For 1-minute Wind data, $M = 720$ corresponds to a 1-day cutoff. The boxcar function in frequency space is
\begin{equation}
    \hat{H}(\omega) = \begin{cases}
        1 & |\omega| \leq \frac{\pi}{M},\\
        0 & |\omega| > \frac{\pi}{M}.
    \end{cases}
\end{equation}

Let $\hat{X}(\omega)$ be the Fourier transform of $x(t)$. By the convolution theorem, multiplication by a box-car filter in frequency domain corresponds to convolution with a sinc function in the time domain. So the low-pass-filtered data becomes
\begin{equation}
    x_\mathrm{low}(t) = \mathcal{F}^{-1}[\hat{H}(\omega) \hat{X}(\omega)] = \int_{-\infty}^\infty h(t') x(t-t') dt' = [h * x] (t), \indent \text{where } h(t) \equiv \frac{\sin(\pi t/M)}{\pi t}.
\label{eq:sincfilter2}
\end{equation}
By construction, $\hat{H}(0) = 1$ so $h(t)$ is normalized. The high-frequency component is therefore
\begin{equation}
    x_\mathrm{high}(t) = x(t) - x_\mathrm{low}(t).
\end{equation}

We now implement the procedure above on discrete, bounded time domain, where the data becomes $\{x_i\}$ with $i = 1, \cdots, N$. First, we need to truncate $h(t)$ to finite domain of length $L$, which must be sufficiently large to suppress energy leakage beyond the Nyquist frequency, but not excessively large to avoid unnecessary data loss at small $i$ (the first filtered data point starts at $i=(L+1)/2$). We set $L \approx 10 M$ with $L$ chosen to be odd so $h$ has well-defined center. Now $h$ resides in the positive index domain with center index $i_0 = (L+1)/2$:
\begin{equation}
    h_i = \begin{cases}
        \frac{1}{M} & i = i_0,\\
        \frac{\sin (\pi (i-i_0) / M)}{\pi (i-i_0)} & 1 \leq i \leq L, i \neq i_0,\\
        0 & i > L.
    \end{cases}
\end{equation}
We further multiply $h$ by a Hamming window to avoid real-space discontinuity and reduce spectral leakage. The final filtering function is
\begin{equation}
    g_i = h_i \left( 0.54 - 0.46 \cos{\left[ \frac{2\pi (i-1)}{L-1}\right]} \right).
\end{equation}

Eq.~\ref{eq:sincfilter2} can now be discretized in bounded domain as
\begin{equation}
    x_{\mathrm{low}, k} = \frac{\sum_{i=1}^L g_i x_{k-i_0 + i}}{\sum_i g_i}, \indent k= i_0, \cdots, N-i_0+1.
\end{equation}
And $x_{\mathrm{high}, k} = x_k - x_{\mathrm{low}, k}$. 

\begin{figure}[h]
\centering
    \includegraphics[angle=0,width=0.5\columnwidth]{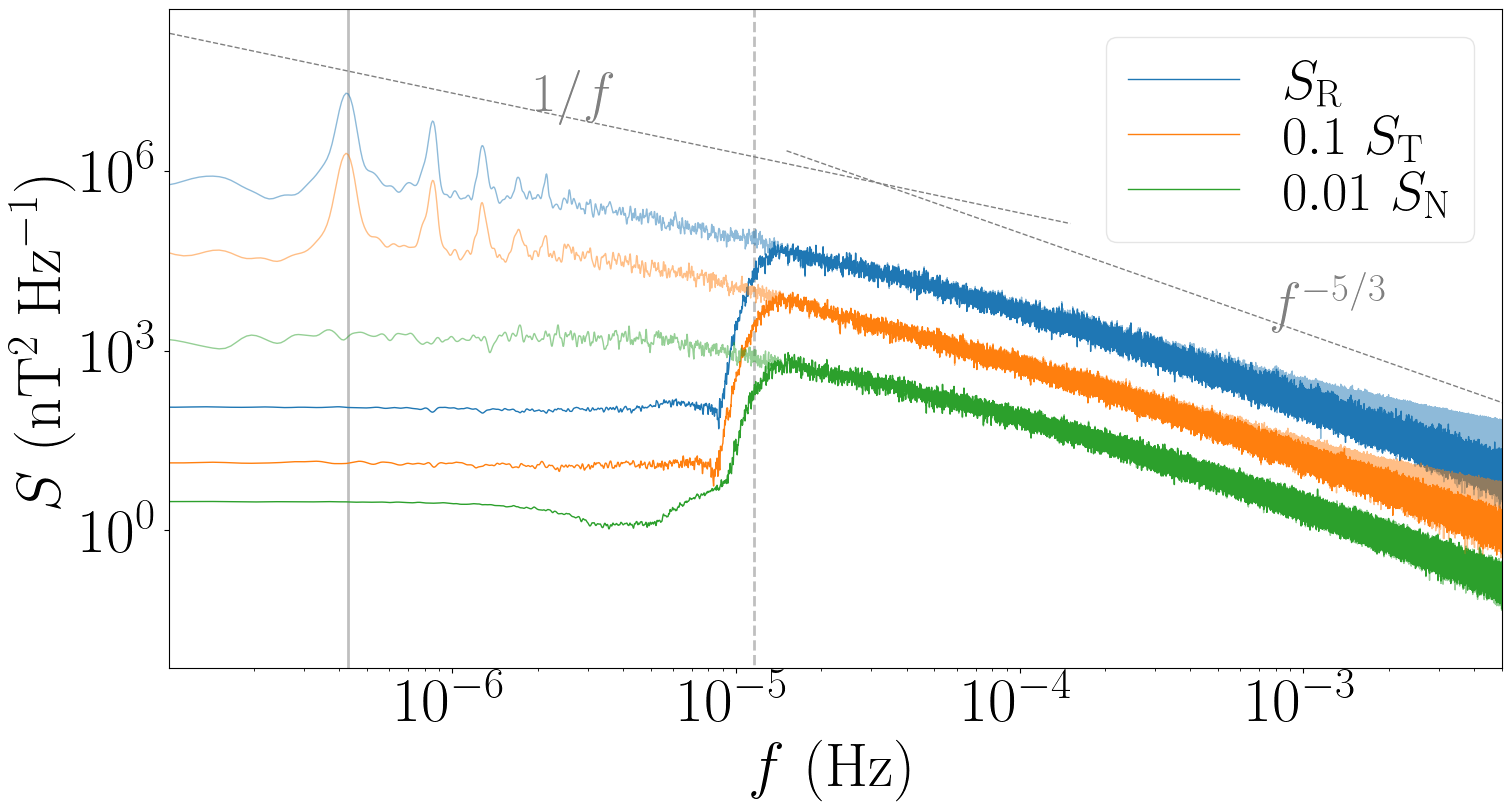}
    \caption{Power spectral densities of magnetic field components before (transparent) and after (solid) 1-day high-pass filtering of Wind data. Vertical gray dashed and solid lines correspond to 1 day and 27 days, respectively.}
\label{fig:highfilter}
\end{figure}

The filtering is applied independently to each magnetic field component. The power spectral densities before and after filtering are compared in Fig.~\ref{fig:highfilter}.

\begin{thebibliography}{}
\expandafter\ifx\csname natexlab\endcsname\relax\def\natexlab#1{#1}\fi
\providecommand{\url}[1]{\href{#1}{#1}}
\providecommand{\dodoi}[1]{doi:~\href{http://doi.org/#1}{\nolinkurl{#1}}}
\providecommand{\doeprint}[1]{\href{http://ascl.net/#1}{\nolinkurl{http://ascl.net/#1}}}
\providecommand{\doarXiv}[1]{\href{https://arxiv.org/abs/#1}{\nolinkurl{https://arxiv.org/abs/#1}}}

\bibitem[{{Anselmet} {et~al.}(1984){Anselmet}, {Gagne}, {Hopfinger}, \& {Antonia}}]{Anselmet84}
{Anselmet}, F., {Gagne}, Y., {Hopfinger}, E.~J., \& {Antonia}, R.~A. 1984, Journal of Fluid Mechanics, 140, 63, \dodoi{10.1017/S0022112084000513}

\bibitem[{Barndorff-Nielsen(1997)}]{BarndorffNielsen97}
Barndorff-Nielsen, O.~E. 1997, Scandinavian Journal of Statistics, 24, 1.
\newblock \url{http://www.jstor.org/stable/4616433}

\bibitem[{{Barndorff-Nielsen} {et~al.}(2004){Barndorff-Nielsen}, {Bl{\ae}sild}, \& {Schmiegel}}]{BarndorffNielsen04}
{Barndorff-Nielsen}, O.~E., {Bl{\ae}sild}, P., \& {Schmiegel}, J. 2004, European Physical Journal B, 41, 345, \dodoi{10.1140/epjb/e2004-00328-1}

\bibitem[{{Benzi} {et~al.}(1991){Benzi}, {Biferale}, {Paladin}, {Vulpiani}, \& {Vergassola}}]{Benzi91}
{Benzi}, R., {Biferale}, L., {Paladin}, G., {Vulpiani}, A., \& {Vergassola}, M. 1991, \prl, 67, 2299, \dodoi{10.1103/PhysRevLett.67.2299}

\bibitem[{Blackman \& Tukey(1958)}]{Blackman58}
Blackman, R.~B., \& Tukey, J.~W. 1958, The Measurement of Power Spectra (Dover)

\bibitem[{{Breech} {et~al.}(2003){Breech}, {Matthaeus}, {Milano}, \& {Smith}}]{Breech03}
{Breech}, B., {Matthaeus}, W.~H., {Milano}, L.~J., \& {Smith}, C.~W. 2003, Journal of Geophysical Research (Space Physics), 108, 1153, \dodoi{10.1029/2002JA009529}

\bibitem[{{Bruno} \& {Carbone}(2013)}]{Bruno13}
{Bruno}, R., \& {Carbone}, V. 2013, Living Reviews in Solar Physics, 10, 2, \dodoi{10.12942/lrsp-2013-2}

\bibitem[{{Castaing} {et~al.}(1990){Castaing}, {Gagne}, \& {Hopfinger}}]{Castaing90}
{Castaing}, B., {Gagne}, Y., \& {Hopfinger}, E.~J. 1990, Physica D Nonlinear Phenomena, 46, 177, \dodoi{10.1016/0167-2789(90)90035-N}

\bibitem[{{Frisch}(1995)}]{Frisch95}
{Frisch}, U. 1995, {Turbulence. The legacy of A.N. Kolmogorov}, \dodoi{10.1017/CBO9781139170666}

\bibitem[{{Frisch} \& {She}(1991)}]{Frisch91}
{Frisch}, U., \& {She}, Z.-S. 1991, Fluid Dynamics Research, 8, 139, \dodoi{10.1016/0169-5983(91)90038-K}

\bibitem[{{Horbury} \& {Balogh}(1997)}]{Horbury97}
{Horbury}, T.~S., \& {Balogh}, A. 1997, Nonlinear Processes in Geophysics, 4, 185, \dodoi{10.5194/npg-4-185-1997}

\bibitem[{{Isaacs} {et~al.}(2015){Isaacs}, {Tessein}, \& {Matthaeus}}]{Isaacs15}
{Isaacs}, J.~J., {Tessein}, J.~A., \& {Matthaeus}, W.~H. 2015, Journal of Geophysical Research (Space Physics), 120, 868, \dodoi{10.1002/2014JA020661}

\bibitem[{{Kailasnath} {et~al.}(1992){Kailasnath}, {Sreenivasan}, \& {Stolovitzky}}]{Kailasnath92}
{Kailasnath}, P., {Sreenivasan}, K.~R., \& {Stolovitzky}, G. 1992, \prl, 68, 2766, \dodoi{10.1103/PhysRevLett.68.2766}

\bibitem[{{Kolmogorov}(1962)}]{Kolmogorov62}
{Kolmogorov}, A.~N. 1962, Journal of Fluid Mechanics, 13, 82, \dodoi{10.1017/S0022112062000518}

\bibitem[{{Koval} {et~al.}(2023){Koval}, {Lepping}, \& {Szabo}}]{Koval23}
{Koval}, A., {Lepping}, R.~P., \& {Szabo}, A. 2023, {Wind Magnetic Field Investigation (MFI) Composite Data in RTN Coordinates}, Heliophysics Digital Resource Library (HDRL) dataset, 2023,  HDRL, \dodoi{10.48322/S1D0-5Q92}

\bibitem[{{Lepping} {et~al.}(1995){Lepping}, {Ac{\~{u}}na}, {Burlaga}, {Farrell}, {Slavin}, {Schatten}, {Mariani}, {Ness}, {Neubauer}, {Whang}, {Byrnes}, {Kennon}, {Panetta}, {Scheifele}, \& {Worley}}]{Lepping95}
{Lepping}, R.~P., {Ac{\~{u}}na}, M.~H., {Burlaga}, L.~F., {et~al.} 1995, \ssr, 71, 207, \dodoi{10.1007/BF00751330}

\bibitem[{{Matthaeus} \& {Goldstein}(1982)}]{Matthaeus82-convergence}
{Matthaeus}, W.~H., \& {Goldstein}, M.~L. 1982, \jgr, 87, 10347, \dodoi{10.1029/JA087iA12p10347}

\bibitem[{Matthaeus {et~al.}(2015{\natexlab{a}})Matthaeus, Wan, Servidio, Greco, Osman, Oughton, \& Dmitruk}]{matthaeus2015intermittency}
Matthaeus, W.~H., Wan, M., Servidio, S., {et~al.} 2015{\natexlab{a}}, Philosophical Transactions of the Royal Society A: Mathematical, Physical and Engineering Sciences, 373, 20140154, \dodoi{10.1098/rsta.2014.0154}

\bibitem[{Matthaeus {et~al.}(2015{\natexlab{b}})Matthaeus, Wan, Servidio, Greco, Osman, Oughton, \& Dmitruk}]{Matthaeus15}
---. 2015{\natexlab{b}}, Philosophical Transactions of the Royal Society A: Mathematical, Physical and Engineering Sciences, 373, 20140154, \dodoi{10.1098/rsta.2014.0154}

\bibitem[{{Novikov}(1971)}]{Novikov71}
{Novikov}, E. 1971, Journal of Applied Mathematics and Mechanics, 35, 231, \dodoi{10.1016/0021-8928(71)90029-3}

\bibitem[{{Obukhov}(1962)}]{Obukhov62}
{Obukhov}, A.~M. 1962, J. Geophys. Res., 67, 3011, \dodoi{10.1029/JZ067i008p03011}

\bibitem[{{Padhye} {et~al.}(2001){Padhye}, {Smith}, \& {Matthaeus}}]{Padhye01}
{Padhye}, N.~S., {Smith}, C.~W., \& {Matthaeus}, W.~H. 2001, \jgr, 106, 18635, \dodoi{10.1029/2000JA000293}

\bibitem[{{Ruiz} {et~al.}(2014){Ruiz}, {Dasso}, {Matthaeus}, \& {Weygand}}]{Ruiz14}
{Ruiz}, M.~E., {Dasso}, S., {Matthaeus}, W.~H., \& {Weygand}, J.~M. 2014, \solphys, 289, 3917, \dodoi{10.1007/s11207-014-0531-9}

\bibitem[{{She} \& {Leveque}(1994)}]{She94}
{She}, Z.-S., \& {Leveque}, E. 1994, \prl, 72, 336, \dodoi{10.1103/PhysRevLett.72.336}

\bibitem[{{Smith} {et~al.}(1998){Smith}, {L'Heureux}, {Ness}, {Acu{\~n}a}, {Burlaga}, \& {Scheifele}}]{Smith98}
{Smith}, C.~W., {L'Heureux}, J., {Ness}, N.~F., {et~al.} 1998, \ssr, 86, 613, \dodoi{10.1023/A:1005092216668}

\bibitem[{{Smith} \& {Ness}(2022)}]{Smith22_l2}
{Smith}, C.~W., \& {Ness}, N.~F. 2022, {ACE Magnetic Field (MAG) Geocentric Solar Ecliptic, GSE, and Geocentric Solar Magnetospheric, GSM, Coordinates, Level 2 (H3), 1 s Data}, Heliophysics Digital Resource Library (HDRL) dataset, 2022,  HDRL, \dodoi{10.48322/7XYH-4Z44}

\bibitem[{{Sorriso-Valvo} {et~al.}(1999){Sorriso-Valvo}, {Carbone}, {Veltri}, {Consolini}, \& {Bruno}}]{SorrisoValvo99}
{Sorriso-Valvo}, L., {Carbone}, V., {Veltri}, P., {Consolini}, G., \& {Bruno}, R. 1999, \grl, 26, 1801, \dodoi{10.1029/1999GL900270}

\bibitem[{{Sorriso-Valvo} {et~al.}(2015){Sorriso-Valvo}, {Marino}, {Lijoi}, {Perri}, \& {Carbone}}]{SorrisoValvo15}
{Sorriso-Valvo}, L., {Marino}, R., {Lijoi}, L., {Perri}, S., \& {Carbone}, V. 2015, \apj, 807, 86, \dodoi{10.1088/0004-637X/807/1/86}

\bibitem[{{Sreenivasan} \& {Antonia}(1997)}]{Sreenivasan97}
{Sreenivasan}, K.~R., \& {Antonia}, R.~A. 1997, Annual Review of Fluid Mechanics, 29, 435, \dodoi{10.1146/annurev.fluid.29.1.435}

\bibitem[{{Wan} {et~al.}(2012{\natexlab{a}}){Wan}, {Osman}, {Matthaeus}, \& {Oughton}}]{Wan12_intermittency}
{Wan}, M., {Osman}, K.~T., {Matthaeus}, W.~H., \& {Oughton}, S. 2012{\natexlab{a}}, \apj, 744, 171, \dodoi{10.1088/0004-637X/744/2/171}

\bibitem[{{Wan} {et~al.}(2009){Wan}, {Oughton}, {Servidio}, \& {Matthaeus}}]{Wan09}
{Wan}, M., {Oughton}, S., {Servidio}, S., \& {Matthaeus}, W.~H. 2009, Physics of Plasmas, 16, 080703, \dodoi{10.1063/1.3206949}

\bibitem[{{Wan} {et~al.}(2010){Wan}, {Oughton}, {Servidio}, \& {Matthaeus}}]{Wan10}
---. 2010, Physics of Plasmas, 17, 082308, \dodoi{10.1063/1.3474957}

\bibitem[{{Wan} {et~al.}(2012{\natexlab{b}}){Wan}, {Oughton}, {Servidio}, \& {Matthaeus}}]{Wan12}
---. 2012{\natexlab{b}}, Journal of Fluid Mechanics, 697, 296, \dodoi{10.1017/jfm.2012.61}

\bibitem[{{Wang} {et~al.}(2024){Wang}, {Matthaeus}, {Chhiber}, {Roy}, {Pradata}, {Pecora}, \& {Yang}}]{Wang24_1overf}
{Wang}, J., {Matthaeus}, W.~H., {Chhiber}, R., {et~al.} 2024, \solphys, 299, 169, \dodoi{10.1007/s11207-024-02401-z}

\bibitem[{{Wang} {et~al.}(2026{\natexlab{a}}){Wang}, {Pecora}, {Chhiber}, {Pradata}, {Adhikari}, \& {Matthaeus}}]{Wang26_superposition}
{Wang}, J., {Pecora}, F., {Chhiber}, R., {et~al.} 2026{\natexlab{a}}, \mnras, 548, stag722, \dodoi{10.1093/mnras/stag722}

\bibitem[{{Wang} {et~al.}(2026{\natexlab{b}}){Wang}, {Pecora}, {Chhiber}, {Roy}, \& {Matthaeus}}]{Wang26_correlation}
{Wang}, J., {Pecora}, F., {Chhiber}, R., {Roy}, S., \& {Matthaeus}, W.~H. 2026{\natexlab{b}}, Proceedings of the National Academy of Science, 123, e2519811122, \dodoi{10.1073/pnas.2519811122}

\end{thebibliography}

\end{CJK*}
\end{document}